# Slope plate of sticky soil granular slope instability based on complex network

Shengdong ZHANG[1]  Shihui YOU*[1]  Longfei CHEN[2]  Xiaofei LIU[2]

1. College of Mechanical and Electrical Engineering, Zaozhuang University 277160;

2. College of Civil Engineering and Mechanics, Xiangtan University 411105

**Abstract:** The particle discrete element simulation of the instability and failure process of the granular slope accumulator model when the metal plate continues downward is obtained, and the two-dimensional total velocity vector of soil particle velocity and slope slip during the instability and failure of the slope accumulator are obtained. Macro-response processes such as removing the angle of the crack surface and the average velocity in the y-direction of the slope top of the slope accumulation body. Construct a normal force chain undirected network model of the slope accumulation body particles under natural accumulation, and study the location of its slip surface, and The results are compared with the experimental results. Finally, the complex network method is used to analyze the topological characteristics of the contact force chain network of the particles on the slope top of the slope accumulation body, and the average degree, clustering coefficient and average shortest path are obtained during the slope instability of the slope accumulation body. The evolutionary rule of the method is used to verify its accuracy in combination with the strength reduction method. The research results show that the average shortest path can provide a more effective early warning of the instability and failure of slope deposits. A complex network theory is used to study the macro response of the slope deposits and its force chain. The interrelationship between the macroscopic structure of the network provides a new mathematical analysis method for the study of slope instability.

## 1 Introduction

Slope instability accidents widely exist in geotechnical engineering, environmental science, transportation and many other fields. Slope will lose stability and cause landslide due to rain erosion, earthquake and impact, which will cause huge economic loss to the society and cause huge life threat to the people. Therefore, the study of slope instability has very important theoretical and scientific significance. Limit equilibrium method and finite element method are commonly used for slope stability analysis, but the limit equilibrium method ignores the internal stress-strain relationship when calculating the sliding force and anti sliding force of each rigid block, and the mechanical properties and stress state of different parts of the slope are not considered in the finite element method calculation.The above two methods are difficult to further describe the discrete failure and movement failure of the slope, and it is difficult to reflect the progressive failure process of the slope [1-3]. The discrete element method, which analysis slop failure mode from meso parameters, and express soil particles in a discrete way, it can intuitively understand the location, width, evolution process and energy transformation of the fission region.Zhou Jian [4] carried out numerical simulation on sandy soil slope and cohesive soil slope by using particle discrete element method, and obtained that the overall failure of sandy soil slope is plastic flow state, there is no obvious crack on the slope, while cohesive soil slope is brittle failure, and instability occurs when deformation exceeds a certain amount.Fakhimi, potyondy, backstron, etc[5-8]. studied the deformation and failure characteristics, macro properties and micro characteristics of rock mass through simulated uniaxial compression test. Based on PFC2D and EDEM software, many scholars such as Yang Bing, Zhang Xiaoxue and Yang Ling studied and discussed the macro response of slope failure process[9-13]. Complex networks can effectively characterize the mesoscopic structure of particles and is conducive to the study of particle size effect. For granular slope

---

*Corresponding author

**E-mail address:** 101434@uzz.edu.cn

accumulation, it is necessary to define the contact relationship between particles from the micro view, and complex network can solve this problem. Through the study of displacement deformation and fracture development of slope accumulation body, complex network can quantitatively analyze the instability phenomenon of slope accumulation body model from the micro contact field.In recent years, many scholars have studied the particle filling field based on the complex network theory, and achieved satisfactory results [14-16]. However, at this stage, the analysis methods of complex network are mostly used to analyze the compression process of dense particle filling samples, and there is little research on the evolution of contact force network in the instability process of granular slope accumulation body from the perspective of complex network, which is in the slope instability. It is a new mathematical analysis method. In this dissertation, the stability of slope accumulation is studied by using the particle discrete element method, A static load is applied to the upper part of the slope accumulation body by the downward pressure of the metal plate to simulate the failure and instability process of the slope accumulation body when there is a load on the upper part of the slope accumulation body.The simulation results are compared with the related experiments. Finally, the complex network theory is introduced to analyze the force chain network on the top of the slope accumulation body of cohesive soil particles. Combined with the strength reduction method, the accuracy and feasibility of the complex network method for studying the slope accumulation body are verified

## 2 Numerical Simulation Method and Network Analysis

### 2.1 Particle Discrete Element Method

Hertz Mindlin with bonding bonding model is used, in which particles are bonded together by a certain size of "viscous bond". When the maximum normal and tangential shear stress is reached, the viscous bond will break. The failure of the viscous bond leads to the failure of the slope.The standard Hertz-Mindlin contact model is used to solve the contact problem before the formation of particle bonding. The ratio of normal force to moment ($F_{n,t}/T_{n,t}$) of particles returns to 0 after the particles are bonded at time $t_{BOND}$, and the values of normal force and moment are updated at each calculation time step by the following equation:

$$\delta F_n = -v_n S_n A \delta t \quad (1)$$

$$\delta F_t = -v_t S_t A \delta t \quad (2)$$

$$\delta M_n = -\omega_n S_t J \delta t \quad (3)$$

$$\delta M_t = -\omega_t S_n \frac{J}{2} \delta t \quad (4)$$

$$A = \pi R_B^2 \quad (5)$$

$$J = \frac{1}{2} \pi R_B^4 \quad (6)$$

Where $A$ is the contact area; $S_n$ and $S_t$ are the normal stiffness and tangential stiffness respectively; $R_B$ is the radius of the "viscous bond"; $\delta_t$ is the time step; $v_n$ and $v_t$ are the normal and tangential velocities of particles; $\omega_n$ and $\omega_t$ are the normal and tangential angular velocities respectively. When the normal and tangential stresses exceed the preset threshold, the bond will break. Therefore, the maximum values of normal and tangential stresses are defined as follows:

$$\sigma_{\max} < \frac{-F_t}{A} + \frac{2M_t}{J} R_B \quad (7)$$

$$\tau_{\max} < \frac{-F_t}{A} + \frac{M_t}{J} R_B \quad (8)$$

## 2.2 Complex network theory

Complex network is a network with some or all properties of self-organization, self similarity, attractor, small world and scale-free. The main parameters of complex networks include degree and degree distribution, degree of agglomeration and its distribution characteristics, shortest distance and its distribution characteristics, and scale distribution of connected groups

### 2.2.1 Degree of node

The degree $k_i$ of node $v_i$ is defined as the number of edges connected with the node. Intuitively, the greater the degree of a node, the more important the node is in a certain sense. The average value of degree $k_i$ of all nodes in the network is called the average degree of the network, which is recorded as $\langle k \rangle$

$$\langle k \rangle = \frac{1}{N}\sum_{i=1}^{N} k_i \quad (9)$$

Where, $N$ is the number of node.

### 2.2.2 Clustering coefficient

The clustering coefficient of complex network measures the clustering degree of a complex network. The larger the clustering coefficient of complex network is, the higher the clustering degree of complex network is, the closer the relationship between nodes is. Generally, if a node $i$ in a network $G$ has $k_i$ edges connected with other nodes, this $k_i$ node is called "neighbor" of node $i$. obviously, there may be at most $k_i(k_i-1)/2$ edges between ki nodes. The ratio of the actual number of edges $E_i$ between $k_i$ nodes and the total possible number of edges is defined as the clustering coefficient $C_i$ of node $i$. The average clustering coefficient of all nodes $i$ in the network is the network The clustering coefficient is expressed by $C$, namely:

$$C_i = \frac{2E_i}{k_i(k_i - 1)} \quad (10)$$

$$C = \frac{1}{N}\sum_{i \in G} C_i \quad (11)$$

### 2.2.3 Average path length

The distance of a node $d_{ij}$ is defined as the number of edges on the shortest path connecting two nodes $i$ and $j$. the average path length $L$ in a complex network is defined as the average distance between any two nodes.

$$L = \frac{1}{\frac{1}{2}N(N-1)} \sum_{i,j \in G, i \neq j} d_{ij} \quad (12)$$

## 3 Numerical simulations

### 3.1 Computational model

The discrete element software is used to simulate the process of slope failure. The discrete element model is shown in Fig.1. The slope model is composed of 1637 uniform particles with a diameter of 3mm. Fixed end constraints are applied to the left, right and bottom surface of the slope. At different positions of the slope top, A, B, C and D ($x_A$=-52.5mm, $x_B$=-62.5mm, $x_C$=-72.5mm, $x_D$=-82.5mm) at a certain speed along the Y axis Load the soil slope (the origin is the center of the model, that is (200 / 2120 / 2)

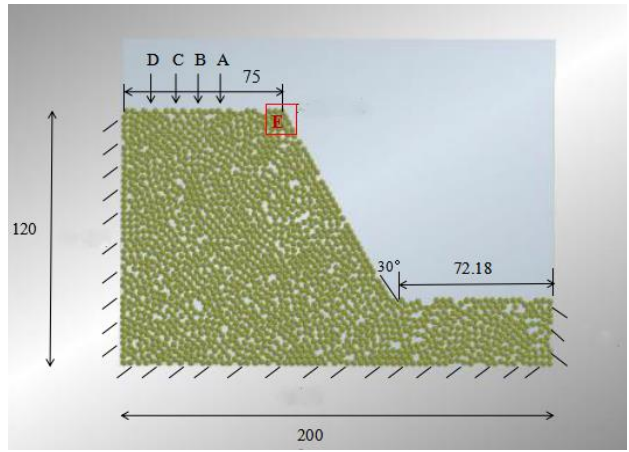

**Fig.1 Slope pile model**

The basic characteristic parameters and bond model parameters of soil granular materials are determined by compression test [17], as listed in Tab.1 and Tab.2.

**Tab.1 Granular material parameters**

| Density kg/m$^3$ | Shear modulus MPa | Stiffness ratio N/m | Poissonby $\nu$ | Static friction coefficient $\mu$ | Rolling friction coefficient $\sigma$ |
|---|---|---|---|---|---|
| 2000 | 469 | 1.5 | 0.35 | 0.5 | 0.001 |

**Tab.2 BPM parameters**

| Normal stiffness GN/m$^3$ | Tangent stiffness GN/m$^3$ | Normal critical pressure MPa | Tangential critical pressure MPa | Glue radius mm |
|---|---|---|---|---|
| 1670 | 667 | 36 | 24 | 0.8 |

### 3.2 Slop failure

When the metal plate focus on position B (xB = - 62.5mm), the velocity vector of soil particles changes with the loading time. At the beginning of loading, the soil on both sides of the metal plate starts to move to the left and right sides due to the extrusion effect. The soil on the left side tends to uplift due to the fixed constraints on the left side, while the soil on the right side moves downward and to the right under the action of deadweight and lower pressure plate due to its proximity to the free end. With the continuous downward action of the metal plate, cracks and shear slip failure appear along the slope direction. When the loading is stopped, most of the soil particles gradually return to the stable state, and only a small amount of soil particles slide down from the slope accumulation under the action of gravity.

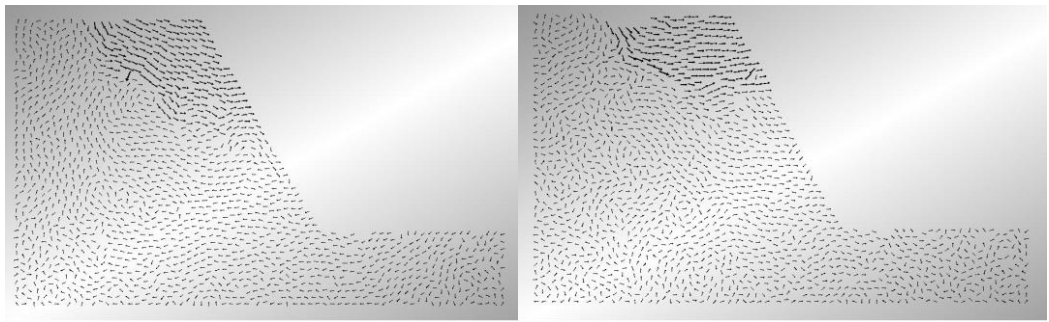

(a) t=0.2s    (b) t=0.6s

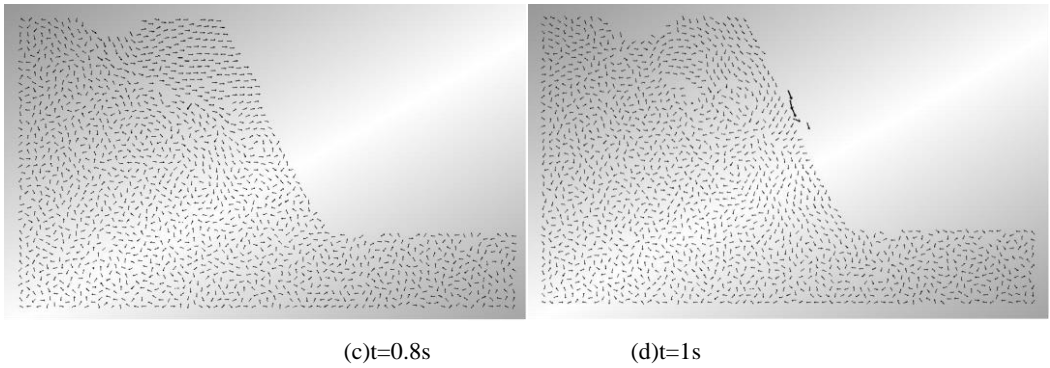

(c)t=0.8s　　　　　　　　(d)t=1s

**Fig.2 Particle velocity vector diagram of slope deposits**

Compared Fig.3 results with the discrete element experiment [18] of the granular accumulation slope. From the simulation results of the change trend of the slope accumulation body and the formation position of the slip crack surface, it can be seen that the instability path is mainly a stepped form composed of steep inclined faults and gently inclined fractures outside the slope, and the instability mode is mainly local shear tension instability mode, The results are basically similar to the experimental results, which verifies that the DEM simulation conforms to the law of landslide evolution. Under the condition of a certain slope strength, the basic form of slope failure is that sliding moves towards the slope direction, and the results are the same as those obtained by Katz [19] and others using DEM

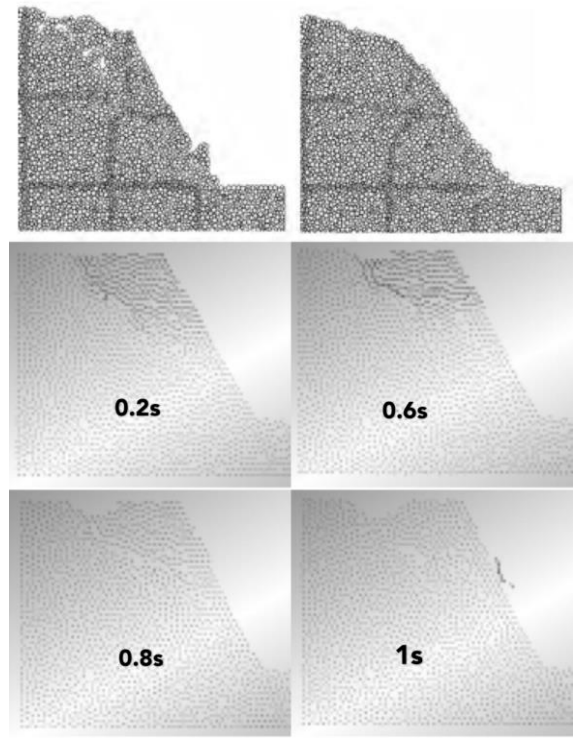

**Fig.3 Comparison of deformation of experimental and analog maps**

### 3.3 Effect of load position

In order to explore the effect of the load position on the slope failure, the discrete element simulation of the slope failure process at different positions of A, B, C, and D is carried out. The results are shown in Figure 3. It changes with the changing position of the metal plate. When loading position A, the angle of slope slipping and cracking surface is 63°, which is slightly larger than the angle of soil slope accumulation; when loading position B, the angle of slope slipping and cracking surface is 40°; When loading position at C, the angle of the slope slipping and cracking surface is 31°; when loading position at D, the slope soil is only squeezed directly below the loading plate, and the soil on both sides of the loading plate remains in a stable state, the tendency of relative motion is small, and there is no slippage and cracking surface.

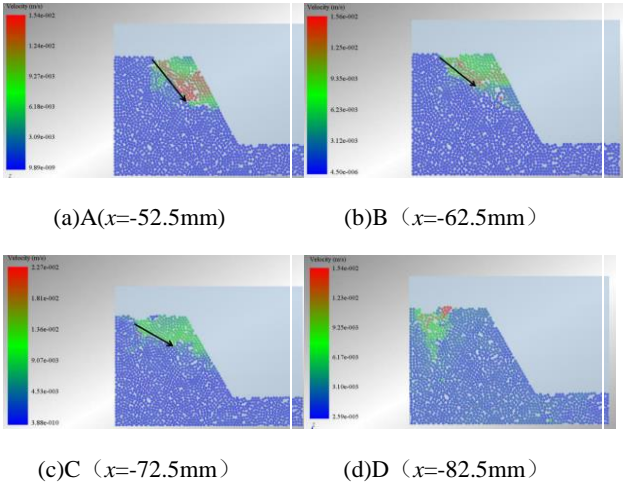

(a)A($x$=-52.5mm)   (b)B ($x$=-62.5mm)

(c)C ($x$=-72.5mm)   (d)D ($x$=-82.5mm)

**Fig.4 Velocity of soil particles in slope piles at different positions**

According to the experiment of soil change law of slope model under local load at the top of slope [20], the maximum horizontal tensile deformation occurs at the upper part of the slope, and the upper soil is more prone to deformation under the local load. Then, according to the constraint direction and particle characteristics of cohesive soil particle accumulation studied in this thesis, the slope top, E point, is selected as the monitoring point. Combined with the simulation results of the previous four groups of slope accumulation instability, the y-direction average velocity of measuring point e at the top of slope is monitored in real time.

Fig.5 is the y-direction average velocity-time curve of the slope top measuring point E at different acting positions. Compare the average velocity-time curve of the slope top measuring point E at the three different acting positions of A, B and C. As the position of the metal plate is farther from the measuring point E on the top of the slope, the peak of the upward movement velocity of the measuring point E appears later. This is because the farther the loading point is, the longer the transfer time of the compression effect of the metal lower plate on the soil; and as the position of the metal plate is farther away from the slope top measuring point E, the lower the peak of the downward movement speed of the measuring point E appears. This is because the farther the metal plate loading point is, the slope top ,the slippage and cracking surface of the soil appears later. From the average speed-time curve of the slope top measuring point E when the acting position is D, it can be seen that when the metal plate acting position is far enough from the slope top measuring point, the compression effect of the metal plate on the soil can be transmitted to the top of the slope, but there will be no slippage and cracking surface.

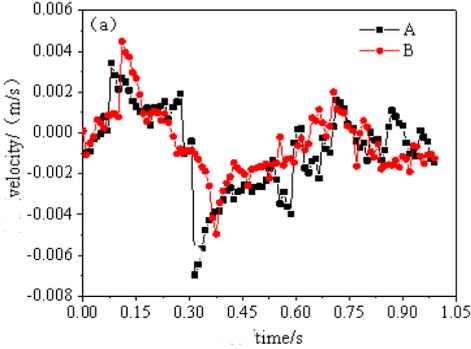

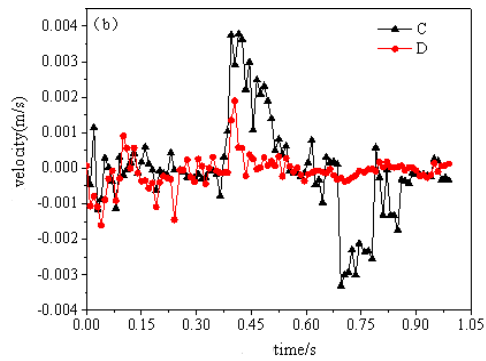

**Fig.5 The average velocity-time curve of the *y*-direction of the Monitoring point E of the slope at different action locations**

## 4 Instability analysis of slope accumulation based on complex network

Using the statistical tools of complex network theory to analyze the network structure of soil particle filling structure, quantify the evolution law of mesoscopic structure of soil particles in slope accumulation under external load, and better characterize the mechanical characteristics of slope accumulation from static accumulation, soil cracking to sliding instability, and the characteristics of slope accumulation body instability process. Pajek software is used to calculate the topological parameters of complex network of soil particles in slope accumulation

### 4.1 Instability Research based on complex network

Figure 6 shows the force chain network under the natural accumulation of soil particles of slope accumulation body. In order to find the location of the slip surface, the contact force network is divided into five regions, as shown in Fig. 6. The average degree of network parameters, average shortest path and clustering coefficient are calculated and analyzed

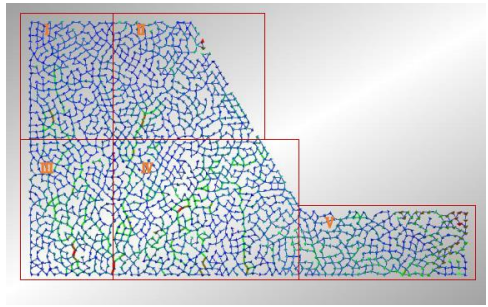

**Fig.6 Force chain network under natural accumulation of soil particles in slope piles**

Fig.7, Fig.8 and Fig.9 show the evolution process of average degree, average shortest path and clustering coefficient of different regions with time when metal plate acts on position B.It can be seen from Figure 7 that at the beginning of loading, a small amount of fracture occurs in the contact between particles in zone II, which leads to a relatively gentle decrease in the average degree of contact force chain network of particles.With the continuous action of load, the contact force is greatly lost, and the average degree of contact force chain network of soil particles on the top of slope accumulation body decreases faster.However, in area of I、III、IV、V, due to the constraint of fixed end, the deformation of extrusion is small, the change of contact force network is small, and the average degree of network is basically unchanged.It can be seen from figure 8 that at the beginning of loading, the average shortest path of particle contact force network in zone II is relatively stable. After loading for a period of time, the average shortest path of particle contact force network on the top of slope accumulation body increases sharply, and finally the average shortest path of the network is relatively stable. It can be seen from Fig. 9 that during the whole

loading process, the clustering coefficient of contact force network in zone II increases gradually with the passage of time, while the change of particle contact force network in zone I、III、IV and V is also small, and the clustering coefficient of network basically does not change.

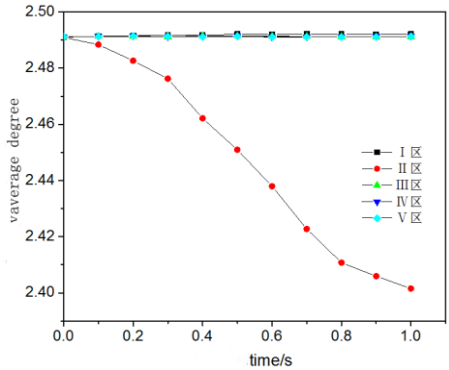

Fig.7 Average degree - time curve of contact force network in different areas

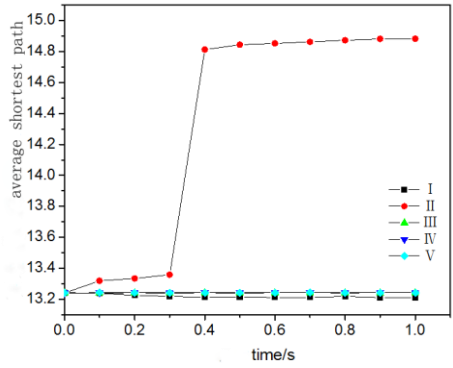

Fig.8 Average shortest path - time curve of contact force network in different regions

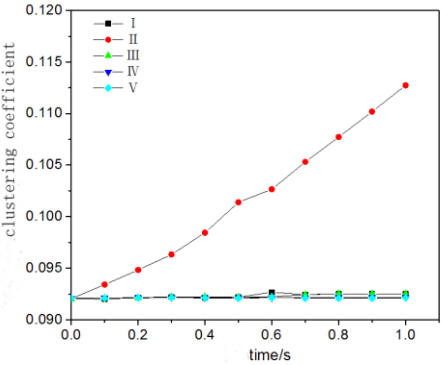

Fig.9 Clustering coefficient - time curves of contact force networks in different regions

By analyzing the evolution trend of network parameters in Fig. 7, FIG. 8 and Fig. 9, it is concluded that the II zone of the network have the largest change and can best reflect the instability process. Therefore, only the II zone of the network are selected for the study.

The average degree time curve of contact force network of soil particles on the top of slope accumulation body during loading is shown in FIG. 10. When the loading position of the metal plate is x = - 52.5mm, x = - 62.5mm, x = - 72.5mm, after loading, due to the extrusion of the metal plate on the soil, the contact between particles appears a small amount of fracture, and the average degree of contact force chain network of particles on the top of slope accumulation body decreases gently; After loading for a period of time, cracks and extension cracks appear in the accumulation body of the slope, and the contact force is greatly missing. The average degree of contact force chain network of soil particles on the top of the slope accumulation body decreases faster; After the formation of sliding crack surface, the slope top of accumulation body forms a relatively stable whole, and the change of contact network and average degree is small. When the loading position of metal plate is x = - 82.5mm, because the loading position of metal plate is far away from the top of slope accumulation, the squeezing effect of soil transferred to the top of slope accumulation is weak, and the number of particle contact force fracture is less.Therefore, the change of contact network of soil particles on the top of slope accumulation is small, and the change of average degree is also small .By comparing the average degree time curves of the contact force network on the top of three slopes, x = - 52.5mm, x = - 62.5mm and x = - 72.5mm, it can be seen that the farther the action position of the metal plate is from the top of the slope accumulation body, the greater the average degree of the final contact force network will be when there is a sliding crack surface.

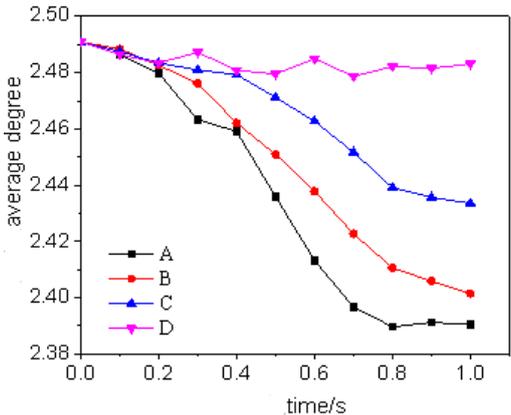

Fig.10 Curve of average degree and time

The clustering coefficient time curve of soil particle contact force network on the top of slope accumulation body during loading is shown in FIG. 11. In the whole loading process, the clustering coefficient increases with time, and the increasing rate increases with the distance between the action position of metal plate and the top of slope accumulation body, and the value of clustering coefficient of final network is also larger. During the loading process, there is a local extrusion effect on the soil particles of the slope accumulation body, which leads to the local aggregation of the contact force network at the top of the slope. The closer the metal plate is to the top of the slope, the greater the deformation of the slope body is, the more obvious the extrusion effect is, and the closer the local agglomeration of the network is.

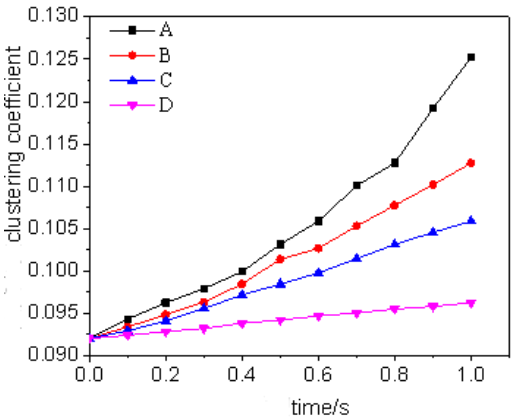

Fig.11 Curve of clustering coefficient and time

The average shortest path time curve of contact force network of soil particles on the top of slope accumulation body during loading is shown in FIG. 12. When the position of the metal plate is $x = -52.5mm$, $x = -62.5mm$, $x = -72.5mm$, the extrusion effect transferred to the top of the slope accumulation is relatively small at the beginning of loading, and the average shortest path of the contact force network of soil particles on the top of the slope accumulation body is relatively stable. After a period of loading, the extrusion effect transferred to the top of the slope accumulation increases, and the accumulation body appears cracks and prolongs The average shortest path of the soil particle contact force network on the top of the slope accumulation body appears a cliff type mutation. After the formation of the sliding crack surface, the slope top of the accumulation body forms a relatively stable whole, and the average shortest path of the network remains relatively stable; When the loading position of the metal plate is x = - 82.5mm, because the loading position of the metal plate is far away from the top of the accumulation body of the slope, the squeezing effect of the soil transferred to the top of the accumulation body of the slope is relatively weak. Therefore, the change of the contact network of the soil particles on the top of the slope accumulation body is small, and the change of the average shortest path is also small; According to the average shortest path time curve of the particle contact force network on the top of the three 72.5mm slopes, we can know that the farther the metal plate acts from the top of the slope accumulation body, the later the value of the average shortest path appears abrupt change, and the smaller the change

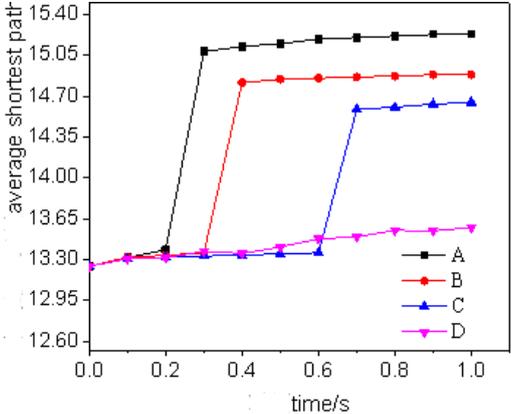

Fig.12 Curve of average shortest path and time

At different loading positions, the time scatter diagram of the peak velocity of particles at the top of slope along the Y direction and the sharp increase of the average shortest path are shown in FIG. 13. The time of the peak velocity of particles on the top of the slope is almost the same as that of the average shortest path. Therefore, we can use the time of the average shortest path to predict the time of the maximum downward velocity of particles at the top of the slope

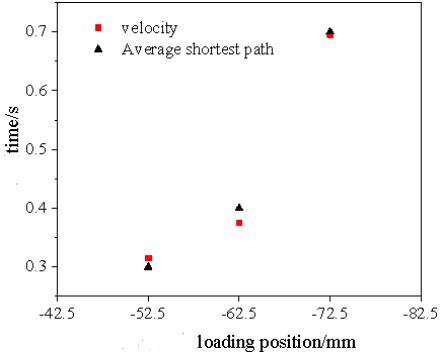

Fig.13 scatter plot of loading position and time

### 4.2 Safety factor based on strength reduction method

The basic principle of strength reduction method[21] is to reduce the strength parameters of materials until failure. Because the discrete element method is used in the simulation, the degradation effect of the tangential bond strength and the normal bond strength is different when considering the micro particles, and the degradation effect of the tangential bond strength is far greater than that of the normal bond strength [22]. Therefore, the normal bond coefficient is defined as 1, and the tangential bond coefficient is reduced. In the particle flow discrete element, the deterioration of the friction coefficient is also far away Less than the shear bond strength. However, in order to improve the accuracy and facilitate the verification of other methods, it is also included in the reduction calculation.

$$c_k = \frac{c}{k} \quad (13)$$

$$\varphi_k = \arctan\left(\frac{\tan\varphi}{k}\right) \quad (14)$$

Where, $c$、$\varphi$ is the true cohesion and internal friction angle of cohesive soil particles; $k$ is the reduction factor; $c_k$、$\varphi_k$ is the cohesion and internal friction angle of reduced cohesive soil particles; The safety factor of slope is the reduction factor of critical state of slope:

$$F_s = k_{critical} \quad (15)$$

There are many methods to judge the instability of the accumulation slope, among which whether the vertical displacement of the top of the slope changes suddenly or not can be used as the criterion of slope instability [23]. For different loading positions (as can be seen from the above analysis, the point loading position is too far away and has little influence, so it is not analyzed). The safety factor of the slope is judged by the trend of the key point velocity time history curve [23]. The reduction factor of three different loading positions corresponding to different loading time is 1.12. When the reduction coefficient is 1.13, the corresponding velocity time histories of the three loading positions do not return to zero at the end of the loading period, but remain a fixed state. It can be seen that when the reduction coefficient is 1.12, the slope is in a critical state. However, the vertical and horizontal velocity time history curves of e tend to zero at 0.2, 0.3 and 0.6 at different loading positions a, B and C, respectively, which is almost the same as the time when the average shortest path suddenly increases, and the error is less than 0.01 s. The accuracy and feasibility of using the average shortest path sudden increase starting point to predict the instability of slope accumulation is verified

## 5 Conclusion

A new analysis method, complex network, is introduced into the study of the topological characteristics of the macro response of slope accumulation and soil particle contact force network. The accuracy and feasibility of this method are verified by combining the experimental model and strength reduction method. The combination of DEM and complex network theory is used to study the instability of slope accumulation. It not only overcomes the defect that the finite element method is only suitable for small deformation, but also solves the high operation cost of traditional discrete element method in calculating the connectivity of rock and soil particles. Compared with the limit equilibrium method, the evolution process of slope deformation is presented.

In the evolution process of contact force chain network of particles on the top of slope accumulation body under the action of metal plate, the change trend of average degree of contact force network with time is roughly consistent. However, the closer the action position of the metal plate is to the top of the slope accumulation body, the decreasing trend of the average contact force network is more obvious. In a certain range, the farther the position of the metal plate is from the top of the slope accumulation, the later the value of the average shortest path increases abruptly, and the smaller the change is. The clustering coefficient increases linearly with the continuous loading. The closer the loading position is to the top of the slope, the faster the growth rate of the clustering coefficient is. When the loading position is far away from the top of the slope, the slope accumulation is relatively stable and the clustering coefficient changes little.

The time of the peak velocity of soil particles at the top of the slope is almost the same as that of the average shortest path, which is consistent with the time when the velocity time history of slope safety factor tends to zero. Therefore, the time when the average shortest path suddenly increases can be used to predict the time when the maximum downward velocity of particles at the top of slope appears, so as to predict the appearance of slip surface time.

The average shortest path of complex network parameters is the most effective one to describe the instability process of slope accumulation. The sudden increase of average shortest path can give early warning of slope accumulation body instability. When the average shortest path of slope particle contact force network increases rapidly, the transmission efficiency of slope accumulation body will also decrease sharply Slip surface appears and instability failure occurs

## Acknowledgement

This work was supported by the Project Funded by Jiangxi Provincial Department of Science and Technology (No. 20192BBEL50028).

## References

[1] T Zhang, H Zheng, C Sun. Global method for stability analysis of anchored slopes[J]. International Journal for Numerical & Analytical Methods in Geomechanics 43(1): 124–137, 2019.

[2] H. Zheng, G. H. Sun, C.G. Li. Cauchy problem of three-dimensional critical slip surfaces of slopes[J].International Journal for Numerical & Analytical Methods in Geomechanics, 35(4): 519 – 527, 2011.

[3] Li Ming-xia, Dong Lian-jie.Analysis on influential factors and deformation characteristics of toppling slope[J].Chinese Journal of Computational Mechanics, 2015(6): 831-837.(in Chinese)

[4] Zhoujian, Wangjiaquan, Zeng Yuan, Zhang Hao. soil slope stability analysis of particle flow simulation[J].Geomechanics, 2009(01).(in Chinese)

[5] Fakhimi A, Carvalho F, Ishida T, et al. Simulation of failure around a circular opening in rock[J]. International Journal of Rock Mechanics & Mining Sciences, 2002, 39(4): 507-515.

[6] Potyondy D O, Cundall P A . A bonded-particle model for rock[J]. International Journal of Rock Mechanics & Mining Sciences, 2004, 41(8): 1329-1364.

[7] Backstrom A, Antikainen J, Backers T, et al. Numerical modelling of uniaxial compressive failure of granite with and without saline porewater[J]. International Journal of Rock Mechanics & Mining Sciences, 2008, 45(7): 1126-1142.

[8] Hsieh Y M, Li H H, Huang T H, et al. Interpretations on how the macroscopic mechanical behavior of sandstone affected by microscopic properties—Revealed by bonded-particle model[J]. Engineering Geology, 2008, 99(1): 1-10.

[9] Yang Bing.Particle dynamics simulation of slope dynamic failure process and collapse range [D]. Tsinghua University, 2011.(in Chinese)

[10] Zhang Xiaoxue. Slope stability analysis based on particle flow simulation [D].Harbin Engineering University, 2015.(in Chinese)

[11] Yang Ling. Discrete element numerical simulation of loess collapse [D]. Xian University of Science and Technology, 2015.(in Chinese)

[12] Feng Chun, LI Shi-hai, SUN Hou-guang, et al. Particle contact meshless method and its application to simulation of slope disaster area[J]. Rock and Soil Mechanics, 2016, 37(12): 3608-3617.(in Chinese)

[13] Zou Yu. Multi-scale analysis of slope stability [D].South China University of Technology, 2017.(in Chinese)

[14] Walker D M, Tordesillas A, Small M, et al. A complex systems analysis of stick-slip dynamics of a laboratory fault[J]. Chaos：An Interdisciplinary Journal of Nonlinear Science，2014，24(1)：013132.

[15] Tordesillas A，Pucilowski S, Walker D M, et al. A Complex Network Analysis of Granular Fabric Evolution in Three-Dimensions[J]. Dynamics of Continuous Discrete & Impulsive Systems， 2012，19(4).

[16] Yi Chenhong，Miao Tiande，Mu Qingsong. Research on Complex Network Method of Particle Media Force Chain[C]. The 2nd National Conference on Computational Mechanics of Particle Materials (CMGM-2014) Proceedings of Conference on Computational Mechanics of Granular Materials. 2014 ：5.(in Chinese)

[17] Quist J, Evertsson C M . Cone crusher modelling and simulation using DEM[J]. Minerals Engineering，2015， 85：92-105.

[18] Wangruhong，Peng Zhouhaiqing, Peng Guoyuan.Dispersion element simulation analysis of mutant instability of piles slope[J].Weapons Equipment Engineering Journal，2018，39(6)： 192-196.(in Chinese)

[19] KATZ O，MORGAN J K，AHARONOV E，et al.Controls on the size and geometry of landslides:insights from discrete element numerical simulations[J]. Geomorphology，2014，220：104-113.

[20] Liao Jingwei，Slope Stability Analysis of Silty Clay Particle Flow Based on Strength Reduction method [D].Chongqing Jiaotong University，2014.(in Chinese)

[21] Songzhanpu，Shibin, Wangyilong, Yan Yifan. Experimental study on distributed optical fiber monitoring of slope deformation by slope cutting [J].Journal of Engineering Geology，2016，24(6)：1110-08.(in Chinese)

[22] Shibutao，Zhangyun, Zhangwei. Material point strength reduction method for slope stability analysis [J]. Journal of Geotechnical Engineering. 2016，09：1678-07.(in Chinese)

[23] Chenyongming，Tengguangliang，Shiyucheng, Qiang Zhengyang. Dispersion element simulation of the instability mechanism of the 109 tunnel slope of Baocheng Railway under the influence of earthquake[J].Journal of Geotechnical Engineering， 2013， S1 0023-10.(in Chinese)